\documentclass[prl,graphicx,amssymb]{revtex4}
  \usepackage{color}
\usepackage{graphicx}
\begin{document}

\title{Comment on ``Which-way information in a nested Mach-Zehnder interferometer''}
\author{ L. Vaidman}
\affiliation{ Raymond and Beverly Sackler School of Physics and Astronomy\\
 Tel-Aviv University, Tel-Aviv 69978, Israel}
\begin{abstract}
 Poto\v{c}ek  and  Ferenczi [Phys. Rev. A {\bf 92}, 023829 (2015)] provided an  analysis of the experimental evidence obtained by  Danan {\it et al.} [Phys. Rev. Lett. {\bf 111}, 240402 (2013)] for the surprising behaviour  of photons passing through an interferometer, in particular, motion on disconnected paths.  Poto\v{c}ek  and  Ferenczi reproduced the results of the experiment,  but when  analyzing its modification, they  claimed  that the reasoning of Danan {\it et al.}, which led to disconnected paths, is erroneous. It is argued here that the criticism of Poto\v{c}ek  and  Ferenczi is unfounded.
\end{abstract}

\maketitle

Poto\v{c}ek  and  Ferenczi (PF) presented a very detailed and correct analysis of the experiment by  Danan {\it et al.} \cite{Danan} in terms of classical optics, significantly extending  the analysis of Saldanha \cite{Salda}. These analyses agree completely with a brief  analysis in the framework of classical optics appearing in \cite{Danan}.
PF correctly state that the classical optics analysis explains all the results without the need for disconnected photon paths. Classical optics does not have a concept of a photon, let alone a post-selected photon, so of course,  there are no photon paths, connected or not.

The novel results of PF are related to a modification of  Danan {\it et al.} experiment. They correctly show that
 the relative heights of the peaks in \cite{Danan} depend on
 the three path lengths, and, in particular, can be tuned such that, without blocking any of the paths,
the trace of only mirror $C$ is present in the output signal. The error of PF is their assertion that the lack of signals from mirrors $A$ and $B$ in their  modification of the experiment is of the same kind as  the lack of signals from mirrors $E$ and $F$ in experiment \cite{Danan}.

Quantum mechanics has a concept of a photon, and there is a clear meaning for the location of the wave packet of a (pre-selected) photon. However, standard quantum mechanics, as classical optics,   has no definition of the location of a pre- and post-selected photon. Danan {\it et al.} experiment tested the paths of photons inside the interferometer according to the {\it definition}, given in \cite{past}, as {\it a place where the photon leaves a weak trace}. The nested interferometer in \cite{Danan} is surprising, mainly because, for a particular setup, the weak trace definition disagrees with the ``common sense''  definition  of Wheeler \cite{Whe}: {\it the photon was present only in the paths through which it could pass}.

Since observing the trace left by  the pre- and post-selected photons  on other systems is an extremely difficult task, in the experiment \cite{Danan},  a degree of freedom of the photons themselves served as a pointer variable for measuring the trace. This was the transversal momentum of the photon which was read off through the positions of the photons on the quad-cell detector.

The two-state vector formalism \cite{AV90} (TSVF) describes a pre-and post-selected photon by a two-state vector.  The TSVF provides a simple way to characterize the weak trace which the pre- and post-selected photon leaves. It vanishes outside the overlap of the forward and the backward evolving states. The strength of the trace in a particular location is characterized by the weak value  of the projection on this location. At the experiment \cite{Danan},  there were non-zero weak values of the projection on the mirrors inside the inner interferometer,  in spite of the fact that, due to the interference, the photons passing through the inner interferometer could not reach the detector. The weak values of the projections on different mirrors were given by Eq. (3) of \cite{Danan}:
\begin{equation}\label{wvs}
({\rm \bf P}_A)_w =({\rm \bf P}_C)_w =1,~~({\rm \bf P}_B)_w =-1,~~({\rm \bf P}_E)_w =({\rm \bf P}_F)_w =0.
\end{equation}
The weak values of the projections show an additional surprising feature. Although there is a trace inside the inner interferometer, there is no trace leading towards and out of it.

Let us consider now the modification of PF in the framework of the TSVF.
To eliminate signals at $A$ and $B$ they suggest to add a phase to path $C$, equal to the Gouy phase $\zeta(z_D)$, without changing anything else. For simplicity, assume that the the detector is far away, so the Gouy phase is $\zeta(z_D)=\frac{\pi}{2}$.
Adding phase $\frac{\pi}{2}$ to path $C$ changes  the two-state vector describing the photon at the intermediate time. Instead of Eq. (1) of \cite{Danan}, it is:
\begin{equation}\label{tsv}
 \nonumber
\langle{\Phi} \vert ~  \vert\Psi\rangle =
{1\over \sqrt 3}\left( \langle A \vert + i\langle B \vert  +   \langle C \vert\right )  ~~ {1\over \sqrt 3}\left( \vert A \rangle +   i\vert B \rangle +  i\vert C \rangle \right ).
\end{equation}
At mirrors $E$ and $F$, the forward and backward evolving states do not overlap, as in the original case. Then, the weak values of the projections on different mirrors are:
\begin{equation}\label{wvs1}
({\rm \bf P}_B)_w =-({\rm \bf P}_A)_w =i,~~({\rm \bf P}_C)_w =1,~~({\rm \bf P}_E)_w =({\rm \bf P}_F)_w =0.
\end{equation}
These values explain the null result in the quad-cell detector at the frequencies of all detectors except $C$, but for very different reasons. At frequencies $f_E$ and $f_F$, the null result is because there is no trace in $E$ and $F$.  At frequencies $f_A$ and $f_B$, the null result is because the quad-cell detector at large distance provides the real part of the weak value of the projections which happens to be zero. The effect of the imaginary part of the weak value on the measuring device is a shift of the conjugate variable, the lateral shift of the beam,  see Eq. (10) of \cite{AV90}. For large $z_D$, it is negligible in comparison to the lateral shift due to the change of the direction of the beam. However, when  the detector is not very far, the effect of the imaginary part is significant. In fact, in the experiment \cite{Danan}, a serious effort was required to keep the phases stable, to avoid imaginary weak values of the projections.

The null signals at frequencies $f_A$ and $f_B$  in the PF modification of the experiment are not because the photons do not leave a trace at mirrors $A$ and $B$, but because the measurement procedure, with a particular distance to the detector, fails to detect it. Changing the position of the detector reveals  the frequencies  $f_A$ and $f_B$. In contrast, keeping destructive interference towards mirror $F$ ensures null signals at  $f_E$ and $f_F$ at any position of the detector.

A subtle point worth repeating \cite{myreply,Sal,ReplySal}, is that when all mirrors  vibrate, none of the signals, including $f_E$ and $f_F$, are exactly zero. And if we postulate that in the experiment there is a zero trace at $E$ and $F$, it is impossible to have signals at  $f_A$ and $f_B$. What allows us to say that there is a trace in $A$ and $B$, but disregard the trace in $E$ and $F$, is that the ratio between the strengths of the traces becomes arbitrary large at the weak limit, while the ratio between the trace in $C$ and the trace in $A$  and $B$, remains constant. If all mirrors vibrate with the same amplitude proportional to a small parameter $\epsilon$, then the traces in $C$, $A$ and $B$ are proportional to $\epsilon$, while the traces in $E$, and $F$ are proportional to $\epsilon^2$.

The photons in the interferometer of Danan {\it et al.} leave a trace which includes a disconnected path. The experiment \cite{Danan} faithfully shows this   (together with a continuous path $C$). In the PF modification of Danan  {\it et al.} experiment the photons leave a trace with a disconnected path too. The null signals at $f_A$ and $f_B$ in  PF's modified experiment arise from an inappropriate method of observing this trace. Therefore, the argument of PF against Danan {\it et al.} conclusions does not hold.

I thank Shimshon Bar-Ad and  Eliahu Cohen for helpful discussions.This work has been supported in part by the Israel Science Foundation  Grant No. 1311/14  and the German-Israeli Foundation  Grant No. I-1275-303.14.

\end{document}